\documentclass[twocolumn,aps,prc,superscriptaddress,showpacs,floatfix]{revtex4-1}
\usepackage{url}
\usepackage{cancel}
\usepackage[colorlinks,linkcolor=blue,citecolor=blue,filecolor=black,urlcolor=blue]{hyperref}
\usepackage{epsfig,graphics}
\usepackage{graphicx}
\usepackage{dcolumn}
\usepackage{bm}
\usepackage[usenames]{color}
\usepackage{amssymb}
\usepackage{amsmath}
\usepackage{multirow}
\usepackage{float}
\usepackage{harpoon}
\usepackage{MnSymbol}
\usepackage{appendix}
\usepackage{color}
\usepackage{hyperref}
\usepackage{cleveref}

\newcommand{\sqrtsnn}{\mbox{$\sqrt{s^{}_{\mathrm{NN}}}$}}

\newcommand{\lr}[1]{\left\langle #1\right\rangle}

\newcommand{\auau}{$^{197}$Au+$^{197}$Au}
\newcommand{\ruru}{$^{96}$Ru+$^{96}$Ru}
\newcommand{\zrzr}{$^{96}$Zr+$^{96}$Zr}
\newcommand{\uu}{$^{238}$U+$^{238}$U}

\newcommand{\avedeltarnp}{\overline{\Delta r_\mathrm{np}}}
\newcommand{\deltarnp}{\Delta r_\mathrm{np}}

\newcommand{\tabincell}[2]{\begin{tabular}{@{}#1@{}}#2\end{tabular}}

\begin{document}
\title{Collision geometry effect on free spectator nucleons in relativistic heavy-ion collisions}
\author{Lu-Meng Liu}
\affiliation{School of Physical Sciences, University of Chinese Academy of Sciences, Beijing 100049, China}
\author{Jun Xu}\email[Correspond to\ ]{xujun@zjlab.org.cn}
\affiliation{School of Physics Science and Engineering, Tongji University, Shanghai 200092, China}
\affiliation{Shanghai Advanced Research Institute, Chinese Academy of Sciences, Shanghai 201210, China}
\affiliation{Shanghai Institute of Applied Physics, Chinese Academy of Sciences, Shanghai 201800, China}
\author{Guang-Xiong Peng}
\affiliation{School of Nuclear Science and Technology, University of Chinese Academy of Sciences, Beijing 100049, China}
\affiliation{Theoretical Physics Center for Science Facilities, Institute of High Energy Physics, Beijing 100049, China}
\affiliation{Synergetic Innovation Center for Quantum Effects $\&$ Applications, Hunan Normal University, Changsha 410081, China}
\date{\today}

\begin{abstract}
Based on the deformed nucleon distributions obtained from the constrained Skyrme-Hartree-Fock-Bogolyubov calculation using different nuclear symmetry energies, we have investigated the effects of the neutron skin and the collision geometry on the yield of free spectator nucleons as well as the yield ratio $N_n/N_p$ of free spectator neutrons to protons in collisions of deformed nuclei at RHIC energies. We found that tip-tip (body-body) collisions with prolate (oblate) nuclei lead to fewest free spectator nucleons, compared to other collision configurations. While the $N_n/N_p$ ratio is sensitive to the average neutron-skin thickness of colliding nuclei and the symmetry energy, it is affected by the polar angular distribution of the neutron skin in different collision configurations. We also found that the collision geometry effect can be as large as 50\% the symmetry energy effect in some collision systems. Due to the particular deformed neutron skin in $^{238}$U and $^{96}$Zr, the symmetry energy effect on the $N_n/N_p$ ratio is enhanced in tip-tip \uu\ collisions and body-body \zrzr\ collisions compared to other collision orientations in the same collision system. Our study may shed light on probing deformed neutron skin by selecting desired configurations in high-energy collisions with deformed nuclei.
\end{abstract}
\maketitle

\section{Introduction}

Distribution of nucleons inside a nucleus is a fundamental probe of nuclear interactions and the nuclear matter equation of state (EOS). The neutron-skin thickness $\Delta r_{\mathrm{np}}$, i.e., generally defined as the difference between the neutron and proton root-mean-square (RMS) radii, is a robust probe of the slope parameter $L$ of the nuclear symmetry energy~\cite{Horowitz:2000xj,Furnstahl:2001un,Todd-Rutel:2005yzo,Centelles:2008vu,Zhang:2013wna,Xu:2020fdc}, characterizing the isospin dependence of the nuclear matter EOS. In the past decades, the $\Delta r_{\mathrm{np}}$ has been measured experimentally through proton~\cite{Zenihiro:2010zz,Terashima:2008zza} and pion~\cite{Friedman:2012pa} scatterings, charge exchange reactions~\cite{Krasznahorkay:1999zz}, coherent pion photoproductions~\cite{Tarbert:2013jze}, and antiproton annihilations~\cite{Klos:2007is,Brown:2007zzc,Trzcinska:2001sy}, etc. More recently, experimental measurement of the $\Delta r_{\mathrm{np}}$ in $^{208}$Pb by parity-violating electron-nucleus scatterings favors a large value of $L$~\cite{Reed:2021nqk}. However, there are a lot of debates on the experimental method~\cite{Corona:2021yfd}, and the resulting large $L$ value is inconsistent with that favored by the electric dipole polarizability~\cite{Piekarewicz:2021jte} or even the $\Delta r_{\mathrm{np}}$ in $^{48}$Ca~\cite{CREX:2022kgg} by a similar measurement method.

Obserables in relativistic heavy-ion collisions are sensitive to the initial condition and thus serve as useful probes of the nucleon distributions in colliding nuclei~\cite{Filip:2009zz,Shou:2014eya,Jia:2021tzt}. In the past few years, significant interest has been induced in this direction by making proposals for the recent isobaric collisions, i.e., \ruru\ and \zrzr\ collisions at $\sqrtsnn=200$ GeV, where various observables at midrapidities are proposed as probes of the neutron-skin thickness in colliding nuclei~\cite{Li:2019kkh,Jia:2021oyt,Xu:2021uar,Jia:2021qyu,Giacalone:2019pca,Bally:2021qys,Jia:2021wbq}. Recently, we proposed that the free spectator neutrons in ultracentral relativistic heavy-ion collisions, which are measurable by zero-degree calorimeters, can be a robust probe of the $\Delta r_{\mathrm{np}}$ in colliding nuclei~\cite{Liu:2022kvz}, free from the uncertainties of modeling the complicated dynamics in the midrapidity region. For a single collision system, we have also proposed that the yield ratio $N_n/N_p$ of free spectator neutrons to protons~\cite{Liu:2022xlm} is a sensitive probe of the $\Delta r_{\mathrm{np}}$ in colliding nuclei, if both spectator neutrons and protons can be measured accurately through dedicated design of the detectors~\cite{Tarafdar:2014oua}.

Density distributions in most nuclei, especially in the vicinity of full shell or subshell, are deformed. The collision dynamics is affected by both the deformation and the collision orientation, among which the tip-tip (with symmetric axis head-on) and body-body (head-on but with symmetric axis parallel) configurations are the most interesting ones. High-energy tip-tip collisions with deformed nuclei, e.g., $^{238}$U, can reach a higher energy density and a larger stopping power, and can thus produce more particles at midrapidities compared to collisions with spherical nuclei, so it is easier for such system to produce the quark-gluon plasma~\cite{Shuryak:1999by,Kharzeev:2000ph,Heinz:2004ir,Miller:2007ri,Kikola:2011zz}. On the other hand, body-body collisions provide the largest overlap region as well as a larger initial eccentricity and thus a larger elliptic flow~\cite{Kolb:2000sd,Hirano:2010jg,Haque:2011aa}. While it is very challenging to select events of special orientations, several promising triggers have been proposed in the literature for tip-tip and body-body collisions~\cite{Nepali:2007an}. Since the neutron skins in deformed nuclei are also deformed, one expects that with proper collision configurations the symmetry energy effect on its probes could be enhanced, similar to what have been observed in intermediate-energy heavy-ion collisions dominated by nucleon degree of freedom~\cite{Xu:2012ys}. Besides the well-known $^{238}$U with a quadrupole deformation of $\beta_2=0.28$~\cite{Moller:2015fba}, an analysis of the ratio of the elliptic flow and the triangular flow in isobaric collisions favors a quadrupole deformation $\beta_2=0.06$ and an octupole deformation $\beta_3=0.20$ for $^{96}$Zr~\cite{Zhang:2021kxj}, and a scaling analysis of the elliptic flow at RHIC energy from colliding nuclei with different quadrupole deformation favors $\beta_2=-0.15$ for $^{197}$Au~\cite{Giacalone:2021udy}. In the present study, we investigate the enhanced effect of the symmetry energy on $N_n/N_p$ in central \zrzr\ and \auau\ collisions at $\sqrtsnn=200$ GeV and \uu\ collisions at $\sqrtsnn=193$ GeV for different collision configurations due to the deformed neutron skins in colliding nuclei.

\section{Theoretical framework}
\label{sec:theory}
In this section, we briefly review the theoretical framework of this study. For more details, we refer the reader to Refs.~\cite{Liu:2022kvz,Liu:2022xlm}.

The spatial distributions of neutrons and protons of initial nuclei are generated based on the energy-density functional from the standard Skyrme-Hartree-Fock (SHF) model, where the 10 parameters in the effective Skyrme interaction can be expressed analytically in terms of 10 macroscopic quantities including the slope parameter $L$ of the symmetry energy~\cite{Chen:2010qx}. The model allows us to vary $L$ while keeping the other parameters fixed at their empirical values~\cite{Chen:2010qx}. As an essential ingredient for the study of open-shell nuclei, the pairing interaction is incorporated when solving the Sch\"ordinger equation in the SHF model, leading to the so-called Skyrme-Hartree-Fock-Bogolyubov (SHFB) model. The pairing interaction between neutrons or protons at $\vec{r}_1$ and $\vec{r}_2$ has the form~\cite{Stoitsov:2012ri}
\begin{equation}
V_{\rm pair}^{(n,p)} = V_0^{(n,p)}   \left(1-\frac{1}{2} \frac{\rho(\vec{r})}{\rho^{}_0} \right)\delta(\vec{r}_1-\vec{r}_2),
\end{equation}
where $V_0^n=-291.5000$ MeVfm$^{-3}$ and $V_0^p=-297.7402$ MeVfm$^{-3}$ are the strength parameters between
neutrons and protons, respectively, $\rho(\vec{r})$ is the local density, and $\rho^{}_0=0.16$ fm$^{-3}$ is the saturation density. Because of the zero-range feature of the pairing interaction, a cutoff $E_{\rm cut}=60$ MeV is introduced in the quasi-particle space. The axially symmetric density distribution of deformed nuclei can then be calculated based on the SHFB calculation by using the cylindrical transformed deformed harmonic oscillator basis~\cite{Stoitsov:2012ri}. With the constrained values of the deformation parameter as listed in Table.\ref{tab:1}, we can get the density distributions in $^{96}$Zr, $^{197}$Au, and $^{238}$U based on the SHFB calculation.

With the above density distribution, we use the Monte-Carlo Glauber model~\cite{Miller:2007ri} to simulate nucleus-nucleus collisions. The nucleon-nucleon (NN) inelastic cross section $\sigma^{}_\mathrm{NN}$ are chosen to be 42 mb at $\sqrtsnn=193$ and 200 GeV. From the above information, the participant nucleons and spectator nucleons are identified. The dynamics of participant matter is completely neglected, since it only affects observables at mid-rapidity region but has no effect on forward and backward regions investigated in the present study. The spectator matter obtained from the Glauber model are further grouped into heavy clusters ($A \geq 4$) and free nucleons based on a minimum spanning tree algorithm. The coalescence parameters are set to be $\Delta r_{\mathrm{max}}=3$~fm and $\Delta p_{\mathrm{max}}=300$ MeV/$c$ as in Ref.~\cite{Li:1997rc}, which have been shown to give the best description of the experimental data of free spectator neutrons in ultracentral \auau\ collisions at $\sqrtsnn=130$ GeV~\cite{Liu:2022kvz}. For spectator nucleons that do not form heavy clusters ($A \geq 4$), they still have chance to coalesce into light clusters with $A \leq 3$, i.e., deuterons, tritons, and $^3$He, and the formation probabilities are calculated according to a Wigner function approach~\cite{Chen:2003ava,Sun:2017ooe}. The total free spectator nucleons are composed of the remaining neutrons and protons that have not coalesced into light clusters, and those from the deexcitation of heavy clusters.

The deexcitation of heavy clusters with $A \geq 4$ are described by the GEMINI model~\cite{Charity:1988zz,Charity:2010wk}, which requires as inputs the angular momentum and the excitation energy of the cluster. The angular momentum of the cluster is calculated by summing those from all nucleons with respective to their center of mass, while the energy of the cluster is calculated based on a simplified SHF energy-density functional which reproduces the same properties of normal nuclear matter as Ref.~\cite{Chen:2010qx}, with the neutron and proton phase-space information obtained from the test-particle method~\cite{Wong:1982zzb,Bertsch:1988ik}. The excitation energy is then calculated by subtracting from the calculated cluster energy the ground-state energy taken from the mass table~\cite{Wang:2021xhn} or an improved liquid-drop model~\cite{Wang:2014qqa}.

\section{Results and discussions}
\label{sec:results}

With the theoretical framework described above, we now discuss the numerical results in the collision systems of \zrzr\ and \auau\ at $\sqrtsnn=200$ GeV, and \uu\ at $\sqrtsnn=193$ GeV. We first give the density distributions of relevant nuclei used in the present study and the corresponding deformed neutron skins. By using the Monte-Carlo Glauber model, we then discuss the collision geometry effect on spectator matter in different collision systems. The discussion will be further focused on experimental observables such as the yield of free spectator nucleons and the yield ratio $N_n/N_p$ of free spectator neutrons to protons in different scenarios.

\subsection{Density distributions of deformed nuclei}

\begin{figure}[htbp!]
	\centering
	\includegraphics[width=1\linewidth]{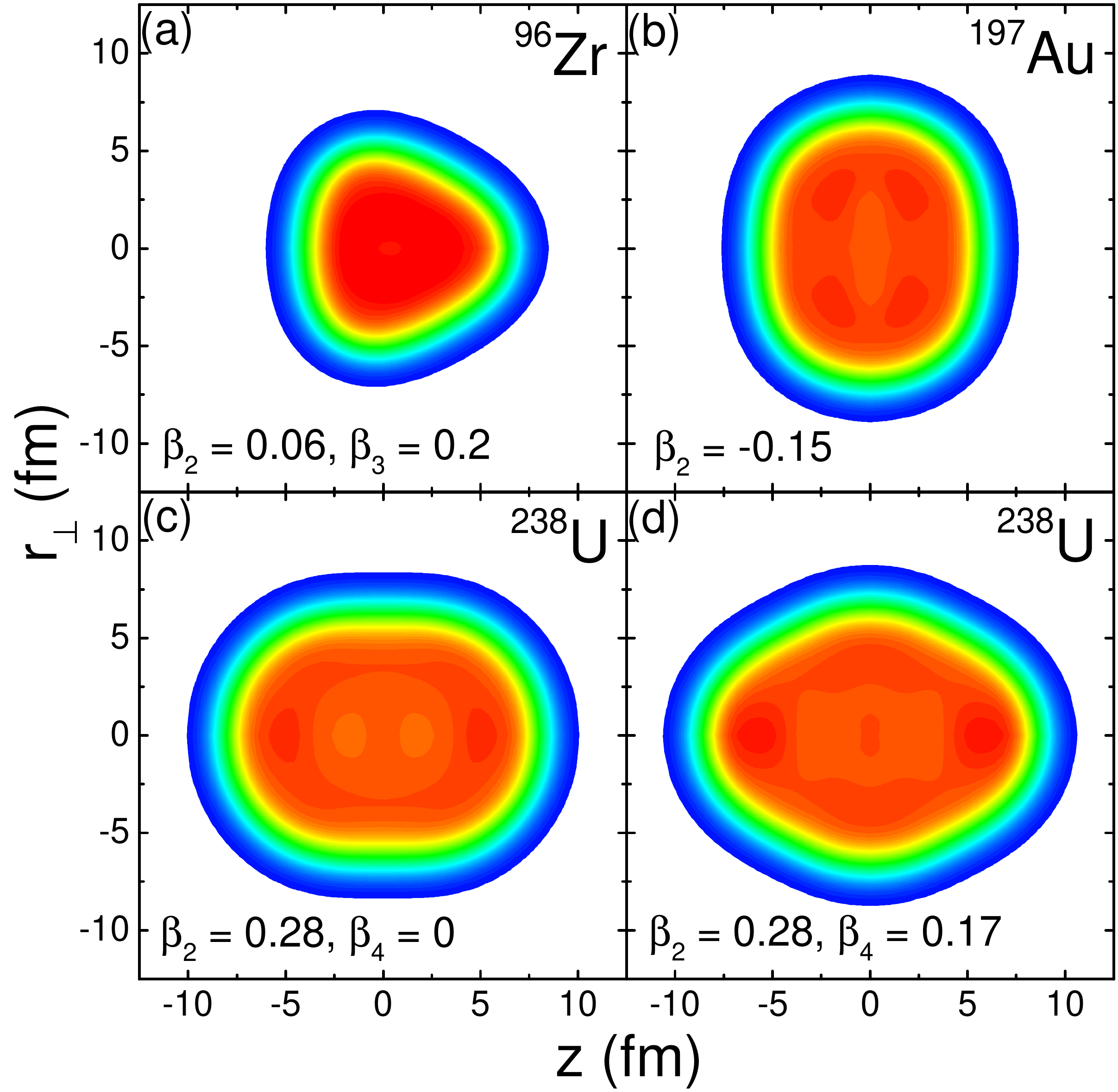}
	\caption{(Color online) Density contours of nucleons in the $r_\perp-z$ plane for $^{96}$Zr with $\beta_2=0.06$ and $\beta_3=0.2$, $^{197}$Au with $\beta_2=-0.15$, and $^{238}$U with $\beta_2=0.28$ and $\beta_4=0$ or 0.17, respectively, from constrained SHFB calculations using $L=120$ MeV.}
	\label{fig:1}
\end{figure}
We display the deformed density contours for nucleons in the $r^{}_\perp-z$ plane of $^{96}$Zr, $^{197}$Au, and $^{238}$U from constrained SHFB calculations for $L=120$ MeV in Fig.~\ref{fig:1}, where the $z$ axis represents the orientation of the symmetric axis and $r^{}_\perp$ is perpendicular to $z$. With the constrained values of $\beta_2$ and $\beta_3$, $^{96}$Zr has a triangular shape with a large octupole deformation and a small quadrupole deformation, while the shape of $^{197}$Au is of an oblate ellipsoid with the symmetric axis $z$ being the short axis. For $^{238}$U, whose shape is of a prolate ellipsoid with the symmetric axis being the long axis from the constrained $\beta_2=0.28$, we investigate two cases with a fixed hexadecapole deformation parameter $\beta_4=0$ or by releasing the constraint on $\beta_4$ which leads to $\beta_4=0.17$ from the SHFB calculation. We note again that the definition of tip-tip (body-body) collisions is the configuration with the $z$ ($r^{}_\perp$) axis head-on.

\begin{table}[!h]
	\normalsize
	\centering
	\caption{Average neutron-skin thicknesses $\avedeltarnp$ for $^{96}$Zr, $^{197}$Au, and $^{238}$U with constrained values of deformation parameters $\beta_{2,3,4}$ using different slope parameters $L$ of the symmetry energy from SHFB calculations.}
	\label{tab:1}
	\renewcommand\arraystretch{1.3}
	\setlength{\tabcolsep}{2mm}
	\begin{tabular}{|c|c|c|c|}
		\hline
		\multirow{2}{*}{Nucleus}    & \multirow{2}{*}{Deformation(s)}&   \multicolumn{2}{c|}{$\avedeltarnp$ (fm)}  \\
		\cline{3-4}
		&                           &   {$L=30$ MeV}  & {$L=120$ MeV } \\
		\hline
		{$^{96}$Zr}                 & $\beta_2$=0.06, $\beta_3$=0.2~\cite{Zhang:2021kxj} & 0.145 & 0.227 \\
		\hline
		$^{197}$Au                  & $\beta_2$=-0.15~\cite{Giacalone:2021udy}  & 0.127 & 0.243\\
		\hline
		\multirow{2}{*}{$^{238}$U } & $\beta_2$=0.28~\cite{Moller:2015fba}, $\beta_4$=0          &  0.156 & 0.291 \\
		& $\beta_2$=0.28~\cite{Moller:2015fba}, $\beta_4$=0.17            &  0.153 & 0.291 \\
		\hline
	\end{tabular}
\end{table}

\begin{figure}[htbp!]
	\centering
	\includegraphics[width=1\linewidth]{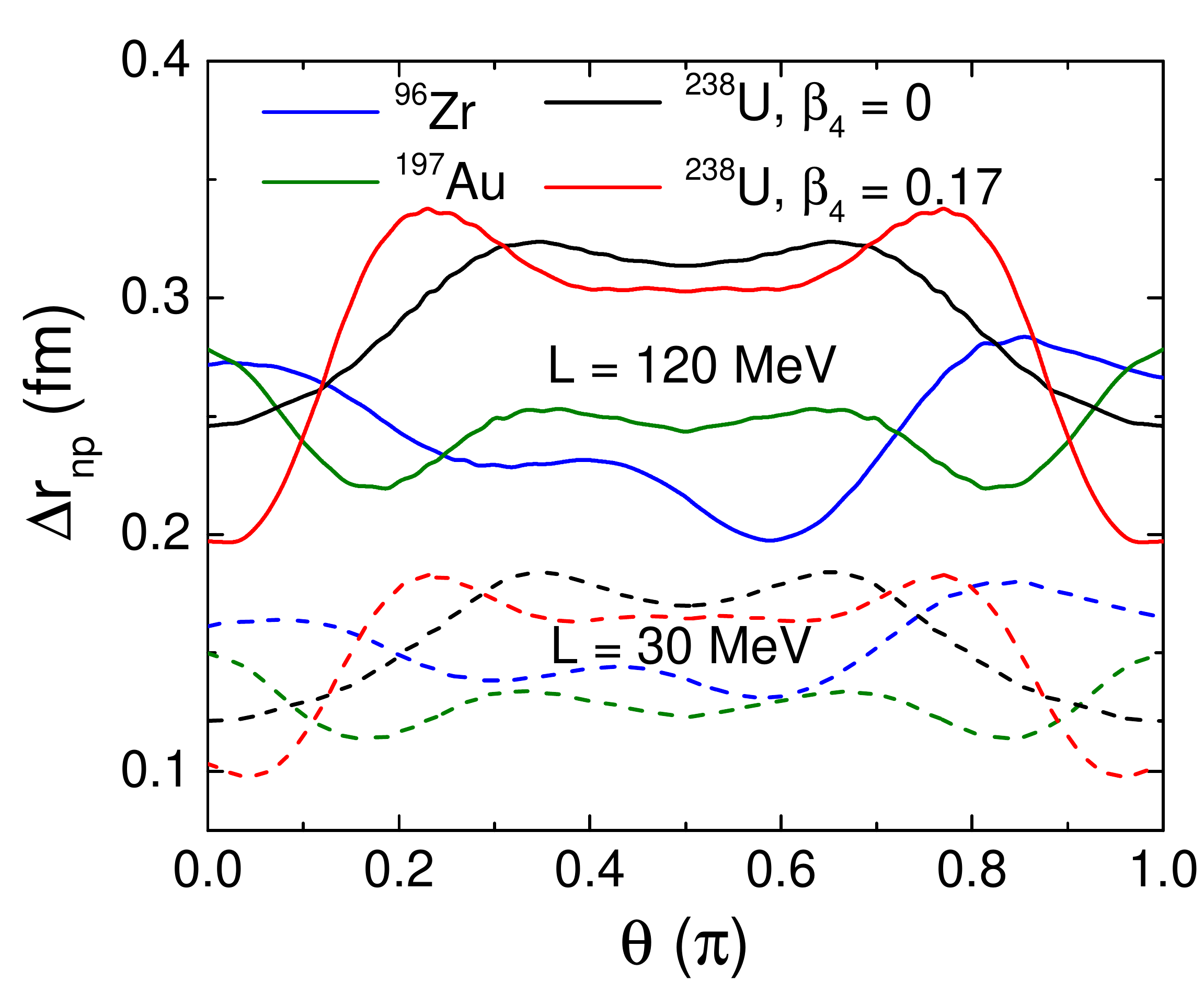}
	\caption{(Color online) Polar angular distribution of the neutron-skin thickness in $^{96}$Zr, $^{197}$Au, and $^{238}$U from constrained SHFB calculations using different slope parameters $L$ of the symmetry energy.}
	\label{fig:2}
\end{figure}

We display the average neutron-skin thickness $\avedeltarnp$ in different scenarios and for different slope parameters $L$ of the symmetry energy in Table \ref{tab:1}. Generally, $\avedeltarnp$ is larger for $L=120$ MeV than for $L=30$ MeV. We note that orientation- averaged charge radii of the corresponding nuclei in different scenarios are consistent with the experimental data~\cite{Angeli:2004kvy} within $1.4\%$. Since both distributions of neutrons and protons in $^{96}$Zr, $^{197}$Au, and $^{238}$U are deformed, in general the neutron skins in these nuclei are deformed as well. Based on the constrained SHFB calculation, the neutron-skin thickness $\Delta r_{\mathrm{np}}$ is a function of the solid angle $\Omega=(\theta,\phi)$, i.e.,
\begin{eqnarray}
	\Delta r_{\mathrm{np}}(\Omega) &=& \sqrt{\lr{r_\mathrm{n}^2(\Omega)}}-\sqrt{\lr{r_\mathrm{p}^2(\Omega)}},
\end{eqnarray}
where
\begin{equation}
	\sqrt{\lr{r_{\tau}^2(\Omega)}} = \left(\frac{\int \rho_{\tau}(r,\Omega) r^4 d{r}}{\int \rho_{\tau}(r,\Omega) r^2 d{r} }\right)^{1/2}
\end{equation}
is the RMS radius for nucleons with isospin index $\tau$ in the direction $\Omega$. In the case of axial symmetry, the solid angular distribution $\Delta r_{\mathrm{np}}(\Omega)$ degenerates to a polar angular distribution $\Delta r_{\mathrm{np}}(\theta)$, which is displayed in Fig.~\ref{fig:2} for different scenarios. One sees that the overall neutron-skin thickness is larger for $L=120$ MeV but has almost the same polar angular distribution compared to that for $L=30$ MeV. The $\theta$ dependence of $\Delta r_{\mathrm{np}}$ is asymmetric for $^{96}$Zr which has a non-zero $\beta_3$, compared with that for $^{197}$Au and $^{238}$U, for which the $\Delta r_{\mathrm{np}}(\theta)$ is symmetric with respect to $\theta=\pi/2$. It is interesting to see that $^{96}$Zr has a larger neutron skin around $\theta \sim 0$ or $\pi$ but a smaller neutron skin around $\theta \sim \pi/2$, while this is opposite to the polar angular distribution of $\Delta r_{\mathrm{np}}$ in $^{238}$U, for which the detailed $\Delta r_{\mathrm{np}}(\theta)$ distribution may also be affected by the value of $\beta_4$. For $^{197}$Au, the angular dependence of $\Delta r_{\mathrm{np}}$ is rather weak. The behavior of $\Delta r_{\mathrm{np}}(\theta)$ may lead to different isospin asymmetries of spectator matter as well as the yield ratios $N_n/N_p$ of spectator neutrons to protons in different collision configurations, to be discussed in the following.

\subsection{Collision geometry effect on spectator matter}

\begin{figure}[htbp!]
	\centering
	\includegraphics[width=1\linewidth]{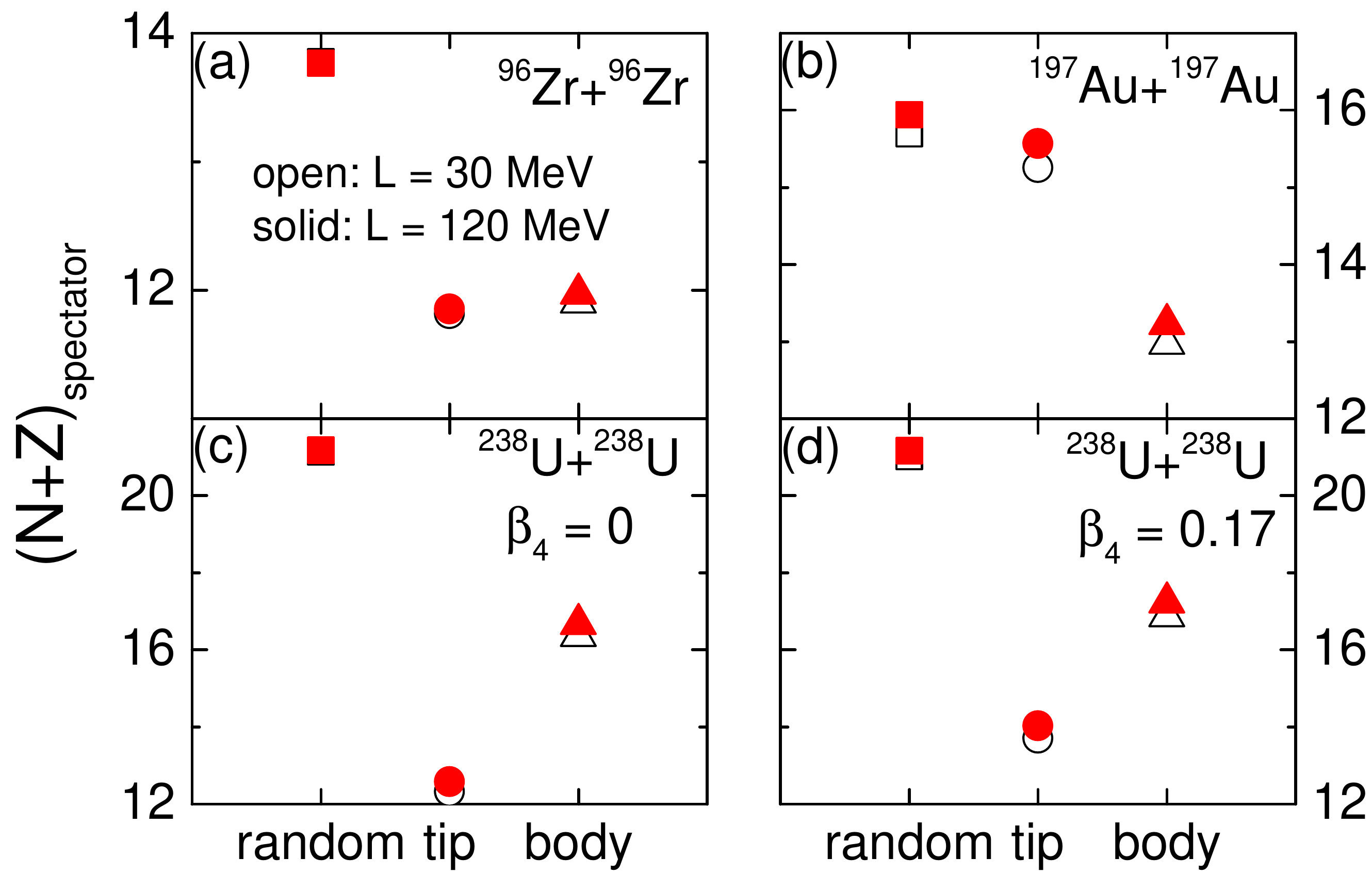}
	\caption{(Color online) Total spectator nucleon numbers in central \zrzr\ and \auau\ collisions at $\sqrtsnn=200$ GeV and \uu\ collisions at $\sqrtsnn=193$ GeV for different collision geometries from the Glauber model with density distributions from constrained SHFB calculations using different slope parameters $L$ of the symmetry energy.}
	\label{fig:3}
\end{figure}
In the present study, we only consider the spectator matter at impact parameter $b=0$ mainly composed of the neutron skin at a particular $\theta$ range depending on the collision geometry, i.e., collisions in random orientations, tip-tip collisions, and body-body collisions. The total spectator nucleon numbers $N+Z$, where $N$ and $Z$ are respectively the total neutron and proton number in the spectator matter, are compared in Fig.~\ref{fig:3} for different scenarios. Generally, the total spectator nucleon number is larger in a heavier collision system. For prolate nuclei with $\beta_2>0$, the total spectator nucleon number is smallest in tip-tip collisions, while it is smallest in body-body collisions for oblate nuclei with $\beta_2<0$. In all cases, the total spectator nucleon number is largest in collisions with random orientations. A larger neutron skin from a larger $L$ slightly increases the total spectator nucleon number.

\begin{figure}[htbp!]
	\centering
	\includegraphics[width=1\linewidth]{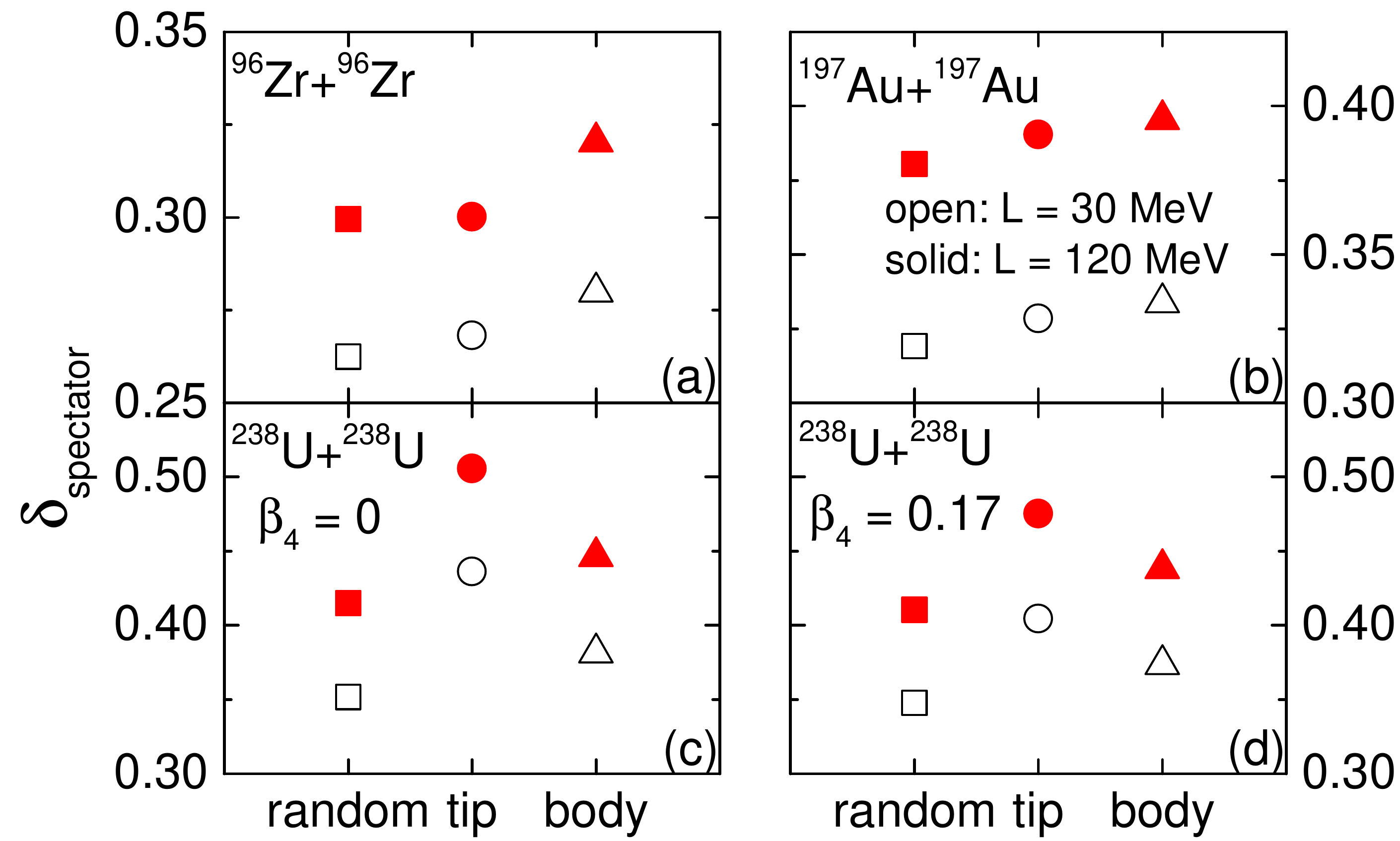}
	\caption{(Color online) Isospin asymmetries of spectator matter in \zrzr\ and \auau\ collisions at $\sqrtsnn=200$ GeV and \uu\ collisions at $\sqrtsnn=193$ GeV for different collision geometries from the Glauber model with density distributions from constrained SHFB calculations using different slope parameters $L$ of the symmetry energy.}
	\label{fig:4}
\end{figure}
The overall isospin asymmetry $\delta_{\rm spectator}=(N-Z)/(N+Z)$ in different collision systems and collision configurations are displayed in Fig.~\ref{fig:4}. As expected, the $\delta_{\rm spectator}$ is larger in collisions by more neutron-rich nuclei. For a given collision system and collision geometry, the spectator matter is more neutron-rich in the case of $L=120$ MeV which leads to a large neutron skin compare with $L=30$ MeV. On the other hand, the $\delta_{\rm spectator}$ also depends on the collision configuration. One sees that the $\delta_{\rm spectator}$ is slightly larger in body-body collisions compared to other collision configurations in \zrzr\ and \auau\ collisions. This is due to the larger neutron skin around $\theta \sim 0$ and $\pi$, which contributes significantly to the spectator matter in body-body collisions. In \uu\ collisions, however, the $\delta_{\rm spectator}$ is larger in tip-tip collisions compared to other collision configurations, as a result of the larger neutron skin around $\theta \sim \pi/2$ than that around $\theta \sim 0$ and $\pi$. Based on the density distribution for $\beta_4=0$, the $\delta_{\rm spectator}$ in tip-tip \uu\ collisions is slightly larger than that for $\beta_4=0.17$, consistent with the slightly larger neutron skin around $\theta \sim \pi/2$ for $\beta_4=0$, as shown in Fig.~\ref{fig:2}.

\subsection{Collision geometry effect on free spectator nucleons}

\begin{figure}[htbp!]
	\centering
	\includegraphics[width=1\linewidth]{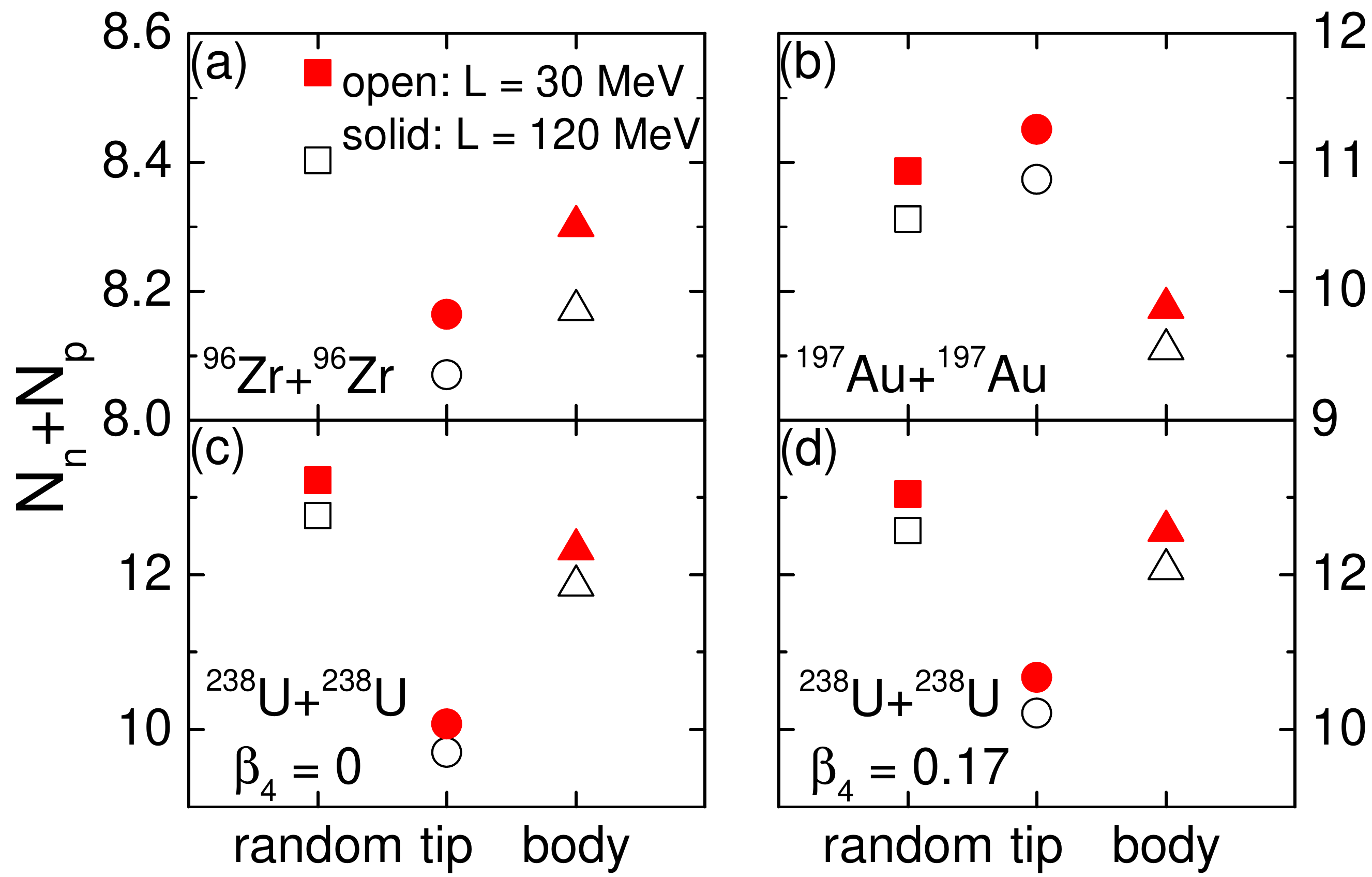}
	\caption{(Color online) Total free spectator nucleon numbers in central \zrzr\ and \auau\ collisions at $\sqrtsnn=200$ GeV and \uu\ collisions at $\sqrtsnn=193$ GeV for different collision geometries from the Glauber model with density distributions from constrained SHFB calculations using different slope parameters $L$ of the symmetry energy.}
	\label{fig:5}
\end{figure}

The free spectator nucleons are composed of the residue ones from direct production that have not coalesced into light clusters and those from the deexcitation of heavy clusters by GEMINI. The numbers of free spectator nucleons in different collision systems and collision configurations are compared in Fig.~\ref{fig:5}. Consistent with the behavior of the total spectator nucleon number as shown in Fig.~\ref{fig:3}, tip-tip collisions lead to fewest free spectator nucleon number in \zrzr\ and \uu\ collisions, while body-body collisions lead to the largest free spectator nucleon number in \auau\ collisions, compared to other collision configurations. A larger $L$ leads to a larger neutron skin and thus more overall free spectator nucleons.

\begin{figure}[htbp!]
	\centering
	\includegraphics[width=1\linewidth]{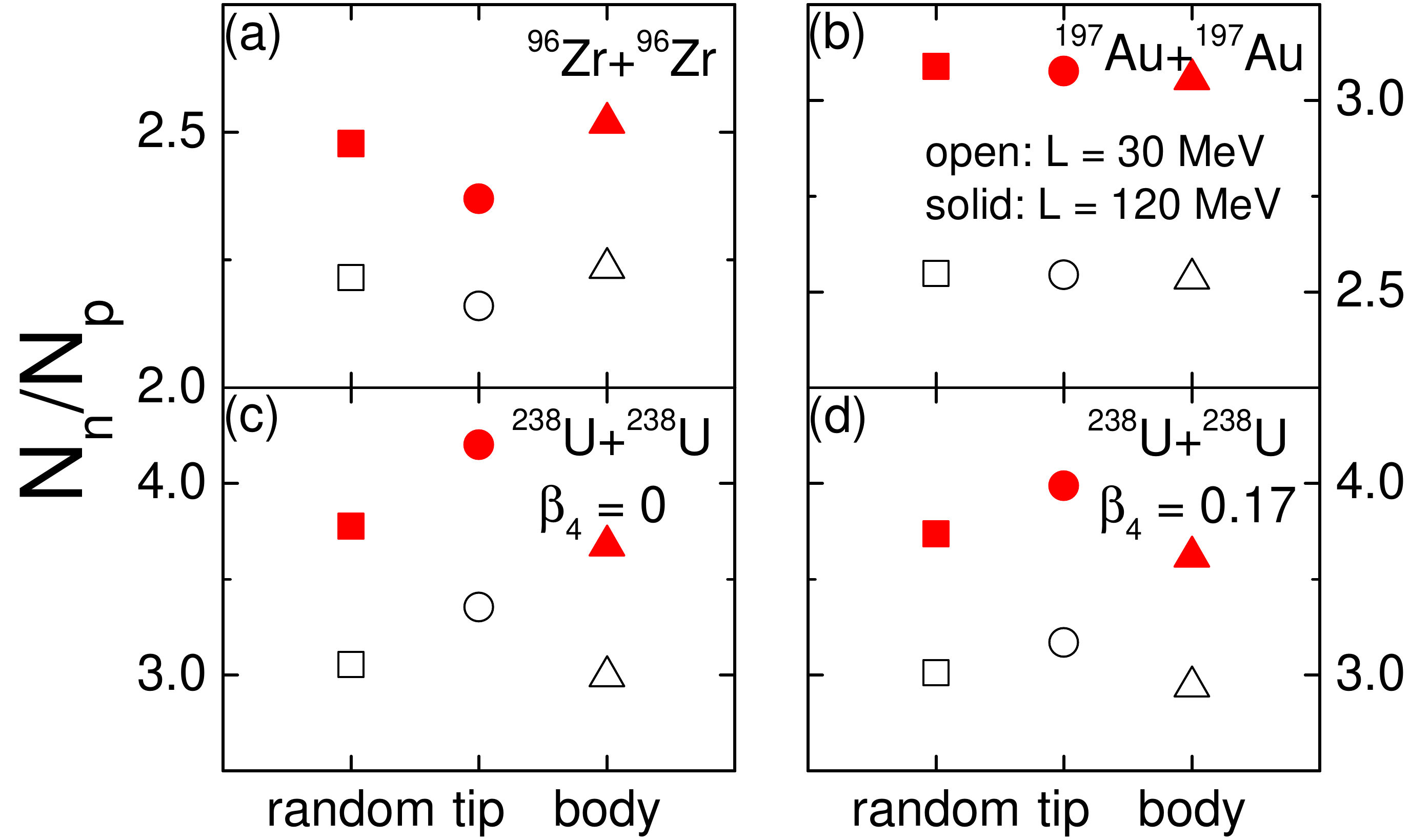}
	\caption{(Color online) Yield ratio $N_n/N_p$ of free spectator neutrons to protons in central \zrzr\ and \auau\ collisions at $\sqrtsnn=200$ GeV and \uu\ collisions at $\sqrtsnn=193$ GeV for different collision geometries from the Glauber model with density distributions from constrained SHFB calculations using different slope parameters $L$ of the symmetry energy.}
	\label{fig:6}
\end{figure}

The yield ratio $N_n/N_p$ of free spectator neutrons to protons was proposed in Ref.~\cite{Liu:2022xlm} as a sensitive probe of the neutron-skin thickness in colliding nuclei and thus the slope parameter $L$ of the symmetry energy. Figures~\ref{fig:6} compares the $N_n/N_p$ ratio in different collision systems and collision configurations. Basically, the behavior of the $N_n/N_p$ ratio is qualitatively consistent with that of the $\delta_{\rm spectator}$, as intuitively expected, since a more neutron-rich spectator matter generally produces more free neutrons than protons. For example, the $N_n/N_p$ ratio is larger in a more neutron-rich collision system, and it is larger in tip-tip collisions compared to other configurations for \uu. On the other hand, the production of free nucleons also depends on the detailed phase-space information of spectator nucleons, which in the present study is consistently given by the constrained SHFB calculation and the Monte-Carlo Glauber model. A slightly smaller $N_n/N_p$ ratio is seen in tip-tip collisions compared with other collision geometries for \zrzr\ collisions, understandable from the deformed neutron skin in Fig.~\ref{fig:2}, but not obviously seen from the behavior of $\delta_{\rm spectator}$ in Fig.~\ref{fig:4}.

\begin{table}[h!]
	\centering
	\caption{Yield ratio $N_n/N_p$ of free spectator neutrons to protons in central \zrzr, \auau, and \uu\ collisions by using different slope parameters $L$ of the symmetry energy in obtaining density distributions of colliding nuclei. The first to third rows of each collision system represent results from collisions with random orientations, tip-tip collisions, and body-body collisions, respectively.}
	\label{tab:2}
	\renewcommand\arraystretch{1.3}
	\setlength{\tabcolsep}{2.5mm}
	\begin{tabular}{|c|c|c|c|}
	\hline
	\multirow{2}{*}{}  &   \multicolumn{2}{c|}{$N_n/N_p$}  & \multirow{2}{*}{$\Delta (N_n/N_p)$} \\
	\cline{2-3}
	&   {$L=30$ MeV}  & {$L=120$ MeV }  & \\
	\hline
	\multirow{3}{*}{\tabincell{c}{\zrzr\\@200 GeV}}              &   2.214$\pm$0.002  &  2.478$\pm$0.003  & 0.263$\pm$0.004\\
	&   2.160$\pm$0.002  &  2.370$\pm$0.003  & 0.210$\pm$0.004\\
	&   2.234$\pm$0.002  &  2.518$\pm$0.003  & 0.284$\pm$0.004\\
	\hline
	\multirow{3}{*}{\tabincell{c}{\auau\\@200 GeV}}              &   2.548$\pm$0.003  &  3.088$\pm$0.003  & 0.540$\pm$0.004\\
	&   2.544$\pm$0.003  &  3.076$\pm$0.003  & 0.532$\pm$0.004\\
	&   2.535$\pm$0.003  &  3.054$\pm$0.003  & 0.520$\pm$0.004\\
	\hline
	\multirow{3}{*}{\tabincell{c}{\uu\\@197 GeV\\$\beta_4=0$}}     &  3.052$\pm$0.003   &  3.774$\pm$0.004  & 0.722$\pm$0.005\\
	&  3.355$\pm$0.004  &  4.202$\pm$0.005  & 0.847$\pm$0.006\\
	&  2.992$\pm$0.003  &  3.676$\pm$0.004  & 0.684$\pm$0.005\\
	\hline
	\multirow{3}{*}{\tabincell{c}{\uu\\@197 GeV\\$\beta_4=0.17$}}  &  3.010$\pm$0.003  &  3.734$\pm$0.004  & 0.725$\pm$0.005\\
	&  3.167$\pm$0.003  &  3.987$\pm$0.004  & 0.818$\pm$0.005\\
	&  2.940$\pm$0.003  &  3.618$\pm$0.004  & 0.677$\pm$0.005\\
	\hline
	\end{tabular}
\end{table}
The effect of the collision geometry on the $N_n/N_p$ ratio as well as its sensitivity to the value of $L$ needs some further discussions. For the ease of discussion, we list the $N_n/N_p$ ratio as well as its difference between calculations using $L=120$ and 30 MeV for different collision systems and collision configurations in Table \ref{tab:2}. One sees that the difference in the $N_n/N_p$ ratio between tip-tip collisions and collisions with random orientations for \zrzr\ and \uu\ collisions can be as large as about $50\%$ that of the symmetry energy effect. By selecting proper collision orientations, one can also get enhanced symmetry energy effect on the $N_n/N_p$ ratio. For example, the difference of the $N_n/N_p$ ratio between $L=120$ and 30 MeV in tip-tip \uu\ (body-body \zrzr) collisions is about 13\% (8\%) larger than that in the case of random orientations. Therefore, the $N_n/N_p$ ratio in tip-tip \uu\ (body-body \zrzr) collisions serves a more sensitive probe of $L$. On the other hand, one sees that the symmetry energy effect is suppressed in tip-tip \zrzr\ collisions. For $^{238}$U nucleus with $\beta_4=0$, the $\deltarnp$ around $\theta \sim \pi/2$ is even larger than that with $\beta_4=0.17$, and this leads to a larger difference in the $N_n/N_p$ ratio between $L=120$ and 30 MeV in tip-tip \uu\ collisions. For \auau\ collisions, however, the $N_n/N_p$ ratio as well as its sensitivity to the value of $L$ is similar in different collision configurations.

\section{Conclusions}

With the deformed nucleon distributions in $^{96}$Zr, $^{197}$Au, and $^{238}$U obtained from the constrained Skyrme-Hartree-Fock-Bogolyubov calculation, we have studied the collision geometry effect on the yield of free spectator nucleons as well as the yield ratio $N_n/N_p$ of spectator neutrons to protons in collisions by these nuclei at top RHIC energy. Tip-tip (body-body) collisions with prolate (oblate) nuclei lead to fewest free spectator nucleons, compared to other collision configurations. While the $N_n/N_p$ ratio is a good probe of the neutron-skin thickness and the slope parameter $L$ of the symmetry energy, we found that the collision geometry effect on the $N_n/N_p$ ratio can be as large as 50\% the symmetry energy effect, due to the deformed neutron skin in colliding nuclei. In addition, although the collision geometry effect in \auau\ collisions is small, we found that the symmetry energy effect is enhanced in tip-tip \uu\ collisions and in body-body \zrzr\ collisions, compared with other collision configurations in the same collision system, as a result of the particular polar angular distribution of the neutron skin in $^{238}$U and $^{96}$Zr. The corresponding $N_n/N_p$ ratio is thus a better probe of $L$ if the collision orientation can be selected in heavy-ion experiments with proper triggers.

Since the deformed nucleon distribution depends on the effective nuclear interaction, the deformed neutron skin is expected to be also sensitive to the effective nuclear interaction, e.g., the nuclear spin-orbit coupling. In this sense, it is promising to study the $N_n/N_p$ ratio in different collision configurations in order to probe the deformed neutron skin, and thus to understand the nuclear structure and nuclear force. Such study is in progress.

\section{Acknowledgments}
We acknowledge helpful discussions with Jiangyong Jia and Chun-Jian Zhang. JX is supported by the National Natural Science Foundation of China under Grant No. 11922514. GXP and LML are supported by the National Natural Science Foundation of China under Grant Nos. 11875052, 11575190, and 11135011.


\bibliography{ref}
\end{document}